\documentclass[10pt, pre,eps,twocolumn,superscriptaddress,showpacs,superscriptaddress]{revtex4-1}
\usepackage{url,ulem}
 \usepackage{fourier}
\usepackage{amssymb,amsmath}
\usepackage[dvipdfmx]{graphicx}
\usepackage{graphicx,color}

\newcommand{\fI}{f_{\rm I}}
\newcommand{\fA}{f_{\rm A}}
\newcommand{\FI}{F_{\rm I}}
\newcommand{\FA}{F_{\rm A}}
\newcommand{\HI}{\mathcal{H}_{\rm I}}
\newcommand{\HA}{\mathcal{H}_{\rm A}}
\newcommand{\dV}{\delta\mathcal{V}}
\newcommand{\ave}[1]{\left\langle #1 \right\rangle}

\newcommand{\aveJI}[1]{\ave{#1}_{\rm I}}
\newcommand{\aveJA}[1]{\ave{#1}_{\rm A}}
\newcommand{\norm}[1]{\left|\left| #1 \right|\right|}

\begin{document}
\title{Landau like theory for universality of critical exponents
  in quasistationary states \\of isolated mean-field systems}
\author{Shun Ogawa}
\email{Shun.Ogawa@cpt.univ-mrs.fr}
\thanks{On leave from Department of Applied Mathematics and Physics, Kyoto University.}
\affiliation{Centre de Phyisque Th\'eorique, UMR7332,
Campus de Luminy, Case 907, 13288 Mareille Cedex 9 ,France}
\author{Yoshiyuki Y. Yamaguchi}
\email{yyama@amp.i.kyoto-u.ac.jp}
\affiliation{
Department of Applied Mathematics and Physics, 
Graduate School of Informatics, Kyoto University, 
606-8501, Kyoto, Japan}
\begin{abstract}
An external force dynamically drives an isolated mean-field Hamiltonian system
to a long-lasting quasistationary state,
whose lifetime increases with population of the system.
For second order phase transitions in quasistationary states,
two non-classical critical exponents have been reported individually
by using a linear and a nonlinear response theories in a toy model.
We provide a simple way to compute the critical exponents all at once,
which is an analog of the Landau theory.
The present theory extends universality class
of the non-classical exponents to spatially periodic one-dimensional systems,
and 
shows that the exponents satisfy a classical scaling relation
inevitably by using a key scaling of momentum.
\end{abstract}
\maketitle 
\section{Introduction}
Universality of critical exponents is one of central issues
in studying phase transitions.
For continuous phase transitions in mean-field systems,
the Landau theory is a powerful tool
to understand the universality and a scaling relation \cite{HN}.
The idea of the Landau theory is,
to construct a pseudo free energy $\mathcal{F}(T,M,h)$ in the form of polynomial by using the Landau expansion, 
\begin{equation}
	\mathcal{F}(T,M, h) = \frac{a(T-T_{\rm c})}{2}M^2 + \frac{b}{4} M^4 + \cdots - hM, 
\end{equation}
where $T$ is temperature,
$T_{\rm c}$ its critical value,
$M$ is the magnetization, 
$h$ the external field, 
and $a$ and $b$ positive constants.
To search the minimal points we consider the condition
\begin{equation}
    \label{eq:Landau0}
    \dfrac{\partial\mathcal{F}}{\partial M}
    = a(T-T_{\rm c})M+ bM^{3} + \cdots - h = 0.
\end{equation}
Let $h$ be sufficiently small and 
$M = m + \delta m$, where $m$ and $\delta m$ represent  respectively 
the spontaneous part and the response to the small external field $h$. 
Then, the equation \eqref{eq:Landau0} 
is divided into the spontaneous part
\begin{equation}
  \label{eq:Landau-spontaneous}
  a(T-T_{\rm c}) m + bm^{3} + \cdots = 0
\end{equation}
and the response part
\begin{equation}
  \label{eq:Landau-response}
  \left[ a(T-T_{\rm c}) + 3m^{2} \right] \delta m + 3 bm(\delta m)^{2}
  + b (\delta m)^{3} + \cdots - h = 0.
\end{equation}
Picking up the first two leading terms in each considering situation,
one can compute the critical exponents
$\beta=1/2,~\gamma_{\pm}=1$ and $\delta=3$ which are defined as
\begin{equation}
  m \propto (T_{\rm c}-T)^{\beta},
  \,\,
  \left.\dfrac{{\rm d}(\delta m)}{{\rm d}h}\right|_{h\to 0} \propto |T-T_{\rm c}|^{-\gamma_{\pm}},
  \,\,
  m \propto h^{1/\delta},
\end{equation}
where $\gamma_{+}$ and $\gamma_{-}$ are defined
in the paramagnetic (Para) high-temperature side
and the ferromagnetic (Ferro) low-temperature side respectively,
and $\delta$ at the critical point.
These exponents satisfy the scaling relation
$\gamma_{\pm}=\beta (\delta-1)$ \cite{HN}.

Before reaching thermal equilibria discussed by the Landau theory,
isolated mean-field Hamiltonian systems are dynamically trapped in
long-lasting quasistationary states (QSSs),
which are vast comparing with thermal equilibria
\cite{AC09, FB10, YL14, AC14}.
The lifetime of a QSS diverges as population of the system \cite{YYY04, JB87},
and it is therefore possible that observable states
are solely QSSs in large population systems.
Elliptical galaxies and the great red spot of Jupiter
are said as examples of QSSs \cite{JB87, AC14}.
The long lifetime naturally induces a question:
Are the critical exponents in the literature of dynamics
the same with of statistical mechanics?
Recently this question is answered negatively.
Dynamics of the mean-field systems is described by the Vlasov equation \cite{BH77},
and a linear \cite{AP12,OY12} and a nonlinear \cite{OY14} response theories
are proposed based on the Vlasov dynamics.
The former gives $\gamma_{+}=2\beta$ but $\gamma_{-}=\beta/2$ \cite{OPY14},
and the latter $\delta=3/2$ \cite{OY14}. 
These exponents satisfies the Widom's scaling relation $\gamma_{-}=\beta (\delta-1)$
irrespective of the value of $\beta$.

However, due to lack of a Landau like theory for QSSs,
which a clue to show the universality of critical exponents,
it has not been clarified how wide the universality class is
and accordingly
whether the scaling relation holds inevitably or accidentally.
There are two obstacles to discuss universality of the critical exponents
in the literature of dynamics.

One is that the exponents are obtained only
in the Hamiltonian mean-field (HMF) model \cite{IK93, MA95}
and partially in the $\alpha$-HMF model \cite{alphaHMF}.
Such systems have particles moving on the unit circle,
and interaction has the first Fourier mode only.
It is not obvious that systems having higher Fourier modes,
for instance the generalized HMF model \cite{TNT12},
also have the same non-classical critical exponents.
Indeed, a non-Hamiltonian model of phase oscillators,
whose continuous version has similar features with the Vlasov equation,
gives $\beta=1/2$ for a single sinusoidal coupling,
but $\beta=1$ in a general coupling \cite{daido-96}.
The other is that computation of the exponent $\delta=3/2$
is independent of $\gamma_{-}$.
The nonlinear response formula gives a self-consistent equation
for the magnetization,
and $\delta$ is obtained by expanding the equation in the Para side,
while $\gamma_{-}$ is defined in the Ferro side.

As the first step to construct the Landau like theory in QSSs,
we provide a simple way to obtain 
an expanded equation of the nonlinear response formula \cite{OY14,YO14},
which is valid both in the Para and the Ferro sides,
and even at the critical point,
for spatially periodic 1D systems with generic interactions.
From the equation, we compute the critical exponents
and show that holding the scaling relation
is inevitable.

This article is organized as follows.
The model and setting are introduced in Sec. \ref{sec:model}. 
In Sec. \ref{sec:expansion},
we first expand the nonlinear response formula \cite{OY14}
around the reference state.
By use of this expansion, in Sec. \ref{sec:expansion-hmf},
we derive the critical exponents $\gamma_{\pm}$
and $\delta$ for the HMF model,
and show that the scaling relation $\gamma_{-} = \beta (\delta - 1)$ 
is inevitable. 
In Sec. \ref{sec:expansion-gen}, this result is generalized
to the Para-Ferro transition in more general models introduced
in Sec. \ref{sec:model}.  
We present summary and discussions in Sec. \ref{sec:summary}. 

\section{Model and setting}
\label{sec:model}
We consider a spatially periodic 1D model
described by the $N$-body Hamiltonian
\begin{equation}
  \label{eq:model}
  H_{N} = \sum_{i=1}^{N} \dfrac{p_{i}^{2}}{2}
  + \dfrac{1}{2N} \sum_{i,j=1}^{N} V(q_{i}-q_{j})
  + \Theta(t) \sum_{i=1}^{N} H_{\rm ext}(q_{i}),
\end{equation}
where $q_{i}\in (-\pi,\pi]$ is the position of $i$-th particle,
and $p_{i}\in\mathbb{R}$ the conjugate momentum.
We assume that the interaction $V(q)$ is even,
and is expanded into the Fourier series
\begin{equation}
  V(q)=-\sum_{k=1}^{K} V_{k}\cos kq,
  \quad V_{k}\neq 0,
\end{equation}
where $K$ is finite.
$H_{\rm ext}$ represents contribution from the external force,
and $\Theta(t)$ is the Heaviside step.
That is, the external force kicks in at $t=0$.
We remark that the following theory can be extended to the external force
which goes to be constant asymptotically instead of the step function.
We also assume that the external part $H_{\rm ext}(q)$
is expanded into the Fourier series
\begin{equation}
  H_{\rm ext}(q) = - \sum_{k=1}^{K} h_{k} \cos kq
\end{equation}
where $h_{k}$ is the conjugate force of $\cos kq$
and is assumed to be small constant.
This model \eqref{eq:model} includes the HMF model \cite{MA95}
by setting $K=1$ and $V_{1}=1$,
and the generalized HMF model \cite{TNT12}
by $K=2$, $V_{1}=\Delta$ and $V_{2}=1-\Delta$.

The corresponding single body effective Hamiltonian is
\begin{equation}
  \begin{split}
    \mathcal{H}[f](q,p,t) &= p^{2}/2 + \mathcal{V}[f] (q,t)
    + \Theta(t) \mathcal{H}_{\rm ext}(q) \\
    \mathcal{V}[f](q,t) &= - \sum_{k=1}^{K} V_{k} (M_{kx}\cos kq + M_{ky}\sin kq) \\
    \mathcal{H}_{\rm ext}(q) & = - \sum_{k=1}^{K} h_{k} \cos kq,
  \end{split}
\end{equation}
where the order parameters are defined as
\begin{equation}
  (M_{kx},M_{ky}) = \int_{\mu} (\cos kq,\sin kq) f(q,p,t) {\rm d}q{\rm d}p,
\end{equation}
with $\mu=(-\pi,\pi]\times\mathbb{R}$.
The single body distribution function $f$ is governed by the Vlasov equation
\begin{equation}
    \partial_{t} f + \{\mathcal{H}[f], f\}=0,
    \quad f(q,p,0) = f_{\rm I}(q,p),
\end{equation}
where the Poisson bracket $\{a, b\}$ is given by 
\begin{equation}
    \{a,b\} = \dfrac{\partial a}{\partial p} \dfrac{\partial b}{\partial q}
    - \dfrac{\partial a}{\partial q} \dfrac{\partial b}{\partial p}.
\end{equation}
We assume $M_{ky}=0~(k=1,\cdots,K)$
and $M_{kx}$ is simply denoted by $M_{k}$,
which is divided into $M_{k}=m_{k}+\delta m_{k}$
where $m_{k}$ and $\delta m_{k}$ are the spontaneous part
and the response to the external field respectively.
In the following, we focus on the phase transition between the Para phase ($m_{1}=\cdots =m_{K}=0$)
and the Ferro phase ($m_{1},\cdots,m_{K}\neq 0$ in general).

We start from a stable stationary state $\fI$
at $t<0$ and exert the external force at $t=0$.
We assume that the external force drives the state
to another stable stationary state $\fA$ asymptotically.
The two stationary states, $\fI$ and $\fA$, give
the Hamiltonians $\HI=\mathcal{H}[\fI]$ and $\HA=\mathcal{H}[\fA]$ respectively,
which differ from each other in general.
Thanks to 1D nature and integrability of $\mathcal{H}_{\rm I/A}$,
angle-action variables $(\theta_{\rm I/A},J_{\rm I/A})$ are available
and $\mathcal{H}_{\rm I/A}$ depends on $J_{\rm I/A}$ only.
We denote the angle averages of an observable $Y$ as
\begin{equation}
  \aveJI{Y} = \dfrac{1}{2\pi} \int_{0}^{2\pi} Y(q,p) {\rm d}\theta_{\rm I},
  \quad
  \aveJA{Y} = \dfrac{1}{2\pi} \int_{0}^{2\pi} Y(q,p) {\rm d}\theta_{\rm A},
\end{equation}
where, for instance, the subscript I of $\aveJI{\cdot}$ represents to take 
the average over each connected iso-$J_{\rm I}$ curve.

\section{Expansion of nonlinear response formula}
\label{sec:expansion}
The nonlinear response theory provides
the asymptotic state $\fA$ which is roughly represented as
$\fA=\aveJA{\fI}$ (see \cite{OY14} for details, and also \cite{XL09}).
Jeans theorem \cite{JHJ15, JB87} states that
$f(q,p)$ is stationary if and only if
it depends on $(q,p)$ solely through the first integrals.
Thus, we may have functions $\FI$ and $\FA$ satisfying
\begin{equation}
  \fI(q,p) = \FI(\HI(q,p)), \quad \fA(q,p)=\FA(\HA(q,p)).
\end{equation}
Our job is to expand $\fA$ around the reference state $\fI$
for computing the small response.

We assume that $\FI$ is given and smooth,
but the form and smoothness of $\FA$
are not obvious due to existence of the average $\aveJA{\cdot}$ \cite{OY14}.
We, therefore, expand $\fA$ by extracting $\fI$
from the averaged form $\ave{\FI(\HI)}_{\rm A}$.The idea is to use the fact
that the bracket $\aveJA{\cdot}$ can be removed
for any function $\psi(\HA)$ as $\aveJA{\psi(\HA)}=\psi(\HA)$,
since the bracket represents the average over an iso-$J_{\rm A}$ curve
while $\HA$ is constant along the curve.
Keeping this fact in mind and 
denoting the order of external force as $O(\mathcal{H}_{\rm ext})=O(h)$,
we expand $\fA=\aveJA{\FI(\HI)}$ as
\begin{equation}
  \label{eq:FIFA}
  \begin{split}
     \aveJA{\FI(\HI)} &=\aveJA{\FI(\HA - \dV)} \\
    &= \FI(\HA) - \FI'(\HA) \aveJA{\dV} \\
    & \quad+ \FI(\HA) \int_{\mu} \FI'(\HA) \dV {\rm d}q{\rm d}p + O(h^{2}),
  \end{split}
\end{equation}
where the asymptotic Hamiltonian $\HA$ is expanded as $\HA = \HI + \dV$
with
\begin{equation}
  \dV
  = - \sum_{k=1}^{K} (V_{k}\delta m_{k}+h_{k})\cos kq
\end{equation}
small. We note that
the third term of the right-hand-side of \eqref{eq:FIFA}
comes from expansion of the normalization factor. 
To understand the third term, we remark that
the normalized $\FI(\HI)$ can be written by
$\FI(\HI)=G(\HI)/\int_{\mu} G(\HI){\rm d}q{\rm d}p$,
where, for instance, the function of energy $G(E)$ is $G(E)=\exp(-E/T)$
if $\fI$ is in canonical thermal equilibrium with temperature $T$.
Using $\HA=\HI+\dV$ again in the first term
of the right-hand-side of \eqref{eq:FIFA}, we have
\begin{equation}
    \label{eq:FIHA}
	\begin{split}
		\FI(\HA)  = &\FI(\HI) + \FI'(\HI) \dV \\
		&- \FI(\HI) 
		\int_{\mu} \FI'(\HI) \dV {\rm d}q {\rm d}p + O(h^2).
	\end{split}
\end{equation}
We may replace $\HA$ with $\HI$ in the third term of \eqref{eq:FIFA}
by omitting $O(h^{2})$ terms,
and the replaced term cancels out with the last term
of \eqref{eq:FIHA}.
Combining them, we have
\begin{equation}
  \label{eq:fAfI0}
   \fA  = \fI + \FI'(\HI) \dV  - \FI'(\HA) \aveJA{\dV} + O(h^{2}). 
  \end{equation}
To clarify the physical interpretation of each term, 
we rewrite it as follows:
\begin{equation}
  \label{eq:fAfI-expand}
  \begin{split}
   \fA & = \fI + \FI'(\HI) \left( \dV - \aveJI{\dV} \right) \\ 
   & + \left[ \FI'(\HI) \aveJI{\dV} - \FI'(\HA) \aveJA{\dV} \right] + O(h^{2}),
   \end{split}
\end{equation}
where the first two terms of the right-hand-side
can be also obtained by the linear response theory \cite{OPY14}, 
and the third one is the main nonlinear effect
of order $o(h)$.
This is the main expansion of this article.

Five remarks for \eqref{eq:fAfI-expand} are in order:
(i) The factors $\aveJI{\dV}$ and $\aveJA{\dV}$
in \eqref{eq:fAfI-expand} are the origin
of the non-classical critical exponents as we will see later.
(ii) It will be shown that the third term is of higher order
than the second term.
(iii) The expansion up to the second term
is consistent with the linear response theory 
based on the Vlasov equation \cite{AP12,OY12}.
Essence of the Vlasov linear response theory 
is to input existence of the Casimir invariants,
$\int_{\mu} s(f) {\rm d}q{\rm d}p$, for $s$ smooth.
Simple algebraic computations reveal that contribution from the term
$-\FI'(\HI)\aveJI{\dV}$ 
keeps the Casimirs
up to the linear order \cite{OPY14}.
(iv) We assumed $\dV$ small but did not assume $m_{k}$ small up to here.
(v) The right-hand-side still depends on $\fA$
through $\HA$, $\dV$ and the average $\aveJA{\cdot}$.
We will introduce self-consistent equations for the order parameters,
whose expansions correspond to the Landau's equation \eqref{eq:Landau0}.

To discuss the critical exponents,
we consider one parameter family of the initial states $\fI$
parameterized by $\tau$ continuously, 
and set the critical point as $\tau=0$.
For instance, $\tau$ is the reduced temperature
$(T-T_{\rm c})/T_{\rm c}$ 
if one considers a family of Boltzmann distributions,
$\FI(\HI)\propto \exp(-\HI/T)$.
Another example is a family of Fermi-Dirac type distributions
where $\FI(\HI)\propto 1/[\exp((\HI-\mu)/T)+1]$.
For a suitably fixed value of $T$, 
this family has a critical point of continuous transition
at $\mu=\mu_{\rm c}$,
and we may set $\tau=\mu-\mu_{\rm c}$ \cite{OPY14}.

\subsection{HMF model case}
\label{sec:expansion-hmf}
It might be instructive to derive the critical exponents
from \eqref{eq:fAfI-expand}
for the HMF model before progressing to the general case.
Let $m_{1}$ be the order parameter in $\fI$
and $m_{1}+\delta m_{1}$ in $\fA$.
The perturbation $\dV$ is $-(\delta m_1 + h) \cos q$ 
in this case.
The self-consistent equation in the asymptotic state
is $m_{1}+\delta m_{1}=\int_{\mu} \fA(q,p) \cos q {\rm d}q{\rm d}p$ and,
by using the main expansion \eqref{eq:fAfI-expand},
it is expanded into
\begin{equation}
  \label{eq:SC-HMF-fI}
  D^{\rm (homo)} m_{1} + B m_{1}^{3} + \cdots = 0
\end{equation}
for the spontaneous part corresponding to \eqref{eq:Landau-spontaneous},
and
\begin{equation}
  \label{eq:SC-HMF-response}
  D (\delta m_{1} +h_{1}) + C (\delta m_{1} +h_{1}) - h_{1} = O(h_{1}^{2})
\end{equation}
for the response part corresponding to \eqref{eq:Landau-response}.
Here,
\begin{equation}
  \label{eq:D-HMF}
  D = 1 + \int_{\mu} \FI'(\HI) \left( \cos q - \aveJI{\cos q} \right) \cos q {\rm d}q{\rm d}p,
\end{equation}
\begin{equation}
  C = \int_{\mu} \left[ \FI'(\HI) \aveJI{\cos q} - \FI'(\HA) \aveJA{\cos q} \right] \cos q {\rm d}q{\rm d}p,
\end{equation}
and $D^{\rm (homo)}$ is defined by forcedly setting $\HI=p^{2}/2$
and $\aveJI{\cos q}=0$ accordingly in \eqref{eq:D-HMF}.
The functional $B$ is obtained by expanding $\FI(\HI)$ with respect to $m_{1}$,
and is assumed to be positive.
The functional $D$ is called the dispersion function,
or the dielectric function in the literature of plasma,
and the state $\fI$ is stable if and only if $D>0$ \cite{SO13}.
The functional $C$ goes to zero in the limit of $h_{1}\to 0$,
in other words ${\rm A}\to {\rm I}$,
and the second term of \eqref{eq:SC-HMF-response} is
of higher order than the first.
Following the spirit of Landau theory,
we compute the critical exponents by picking up
the first two leading terms.

The dispersion function for homogeneous state, $D^{\rm (homo)}$,
is positive (resp. negative) in the Para (resp. Ferro) sides,
and spontaneous magnetization in the Ferro side is
$m_{1}\propto \sqrt{-D^{\rm (homo)}}$.
In general, we may expect $|D^{\rm (homo)}|\propto \tau$
around $\tau=0$ and hence $\beta=1/2$.

In the response part, 
the first two leading terms make
\begin{equation}
  D ( \delta m_{1} +h_{1}) - h_{1} = 0,
\end{equation}
and the critical exponents $\gamma_{\pm}$ are determined
by the convergent speed of $D$ to zero.
To discuss $D$ in the two phases separately,
we denote $D$ in the Para and the Ferro sides
by $D^{\rm (Para)}$ and $D^{\rm (Ferro)}$ respectively.
In the Para side, $D^{\rm (Para)}=D^{\rm (homo)}$
and immediately $\gamma_{+}=2\beta$.
In the Ferro side, we have non-zero $\aveJI{\cos q}$,
and this factor makes the convergence slower.
This slow convergence is observed by 
introducing a new variable 
\begin{equation}
  \label{eq:kappa}
  \kappa = \sqrt{\dfrac{\HI-\HI(0,0)}{\Delta\HI}},
  \quad
  \Delta\HI= \HI(\pi,0) - \HI(0,0) = 2m_{1},
\end{equation}
where $\kappa=0$ at the energy minimum point, the origin,
and $\kappa=1$ on the separatrix.
	The system with the effective Hamiltonian $\mathcal{H}_{\rm I}$ 
	has two fixed points: One is $(0, 0)$ which is the center, 
	and the other is the saddle $(\pi, 0)$ which is identical to $(-\pi,0)$. 
	The separatrix $\{(q,p)|\kappa = 1\}$ is the iso-energy set 
	which consists of stable and unstable manifolds of the saddle,
        and connects the two (identical) saddles.
        The separatrix width to momentum direction is
        of $O(\sqrt{\Delta\HI})=O(\sqrt{m_{1}})$
        from the definition of $\Delta\HI$,
        and plays an important role to obtain
        the critical exponent $\gamma_{-}$.
        To observe it, we first note that \cite{OPY14}
        \begin{equation}
          1 + \int_{\mu} \FI'(\HI)\cos^{2}q {\rm d}q{\rm d}p
          = O(|D^{\rm (homo)}|) = O(m_{1}^{2}).
        \end{equation}
        Next, we focus on the remaining part of $D$, namely
        $\int_{\mu} \FI'(\HI)\ave{\cos q}_{\rm I}\cos q{\rm d}q{\rm d}p$.
        This term goes to zero as $\tau\to 0$
        since $\ave{\cos q}_{\rm I}\to 0$,
        and hence inhomogeneous nature,
        in other words non-zero separatrix width,
        controls the convergent speed of the term to zero.
        We hence extract the convergent speed
        from the integral by scaling the separatrix width to a constant.
        Remembering that $\kappa=1$ represents the separatrix,
        we change the variable from $p$ to $\kappa$ and
        ${\rm d}q{\rm d}p\propto \sqrt{\Delta\HI}{\rm d}q{\rm d}\kappa$.
        Consequently, we have the estimation of
        $D^{\rm (Ferro)}=O(\sqrt{\Delta\HI})= O(\sqrt{m_{1}})$,
        since the scaled integral does not vanish at the critical point
        \cite{OPY14} and
        the term $\int_{\mu} \FI'(\HI)\ave{\cos q}_{\rm I}\cos q{\rm d}q{\rm d}p$ dominates the other.
The exponent $\gamma_{-}$ is, therefore, $\gamma_{-}=\beta/2$.
We stress that the crucial scaling of this estimation is
$O(p)=O(\kappa\sqrt{\Delta\HI})$ in the definition \eqref{eq:kappa}
to scale the separatrix width to a constant.

At the critical point, the dispersion function $D$ vanishes,
and the first two leading terms make
\begin{equation}
  C(\delta m_{1}+h_{1}) - h_{1} = 0.
\end{equation}
Using $\aveJI{\cos q}=0$ and the same variable transform
from $p$ to $\kappa$ as \eqref{eq:kappa}
with replacing $\HI$ with $\HA$,
we can estimate $C$ as $C\propto\sqrt{\delta m_{1}+h_{1}}$.
Thus, the critical exponent is $\delta=3/2$.

With the aid of above understanding,
we reveal that the scaling relation is inevitable
by generalizing the key scaling as
$O(p)=O(\kappa (\Delta\mathcal{H}_{\rm I/A})^{x})$
with $0<x<1$.
The condition $x<1$ ensures that
the discussed terms are larger than the omitted $O(h_{1}^{2})$.
The same computations with the HMF case give $\gamma_{-}=\beta x$
and $\delta=1+x$,
which satisfy the scaling relation $\gamma_{-}=\beta(\delta-1)$.
Thus, averaged terms of $\aveJI{\dV}$ and $\aveJA{\dV}$,
which appears in the Vlasov (non)linear response theory,
induces the non-classical critical exponents
and the scaling relation inevitably.

\subsection{General case}
\label{sec:expansion-gen}
Let us come back to the general case.
Let $m=(m_{1},\cdots,m_{K})$ be the spontaneous order parameter vector in $\fI$,
and $\delta m=(\delta m_{1},\cdots,\delta m_{K})$ the response
to the external force $h=(h_{1},\cdots,h_{K})$.
As the HMF case, substituting the main expansion \eqref{eq:fAfI-expand}
into the self-consistent equation
\begin{equation}
  m_{k}+\delta m_{k} = \int_{\mu} \fA \cos kq ~ {\rm d}q{\rm d}p,
\end{equation}
we have
\begin{equation}
  \label{eq:SC-general-spontaneous}
  D_{kk}^{\rm (homo)} m_{k} - \varphi_{k}(m) = 0
\end{equation}
for the spontaneous part, and
\begin{equation}
  \label{eq:SC-general-response}
  D ( \Lambda \delta m + h )
  + \Lambda C (\Lambda \delta m + h )
  - h = O(h^{2})
\end{equation}
for the response part.
Here $\Lambda={\rm diag}(V_{1},\cdots,V_{K})$,
$D$ and $C$ are now matrices of size $K\times K$ with the $(k,l)$-elements
\begin{equation}
  \label{eq:D}
  D_{kl} = \delta_{kl}
  + V_{k} \int_{\mu} \FI'(\HI) \cos kq \left( \cos lq - \aveJI{\cos lq} \right) {\rm d}q{\rm d}p
\end{equation}
\begin{equation}
  \label{eq:C}
  C_{kl} = \int_{\mu} \cos kq \left( \FI'(\HI) \aveJI{\cos lq} - \FI'(\HA) \aveJA{\cos lq} \right) {\rm d}q{\rm d}p,\end{equation}
and $D^{\rm (homo)}$ is defined from $D$ as the HMF case,
that is,
\begin{equation}
  (D^{\rm (homo)})_{kl} = \delta_{kl} \left[
    1 + \pi V_{k} \int_{\mu} \FI'(p^{2}/2) {\rm d}p \right].
\end{equation}
The functions $\varphi_{k}(m)$ are polynomials
consisting of monomials whose degrees are more than $1$.
The second term of \eqref{eq:SC-general-response}
is of higher order than the first again.

It might be worth noting the concrete forms of matrix $D$
both in statistical mechanics and in the Vlasov dynamics
by setting the initial state as the canonical equilibrium,
$\FI(\HI)=F_{\rm eq}(\HI)\propto\exp(-\HI/T)$, which implies $\FI'=-\FI/T$.
Let us denote the average over $F_{\rm eq}(\HI)$
by $\ave{\cdot}_{\rm eq}$.
From \eqref{eq:D} the Vlasov dynamics gives
\begin{equation}
  D_{kl} = \delta_{kl} - \dfrac{V_{k}}{T}
  \left( \ave{\cos kq\cos lq}_{\rm eq} - \ave{\cos kq\aveJI{\cos lq}}_{\rm eq} \right).
\end{equation}
On the other hand, expanding $\fA\propto\exp(-\HA/T)$,
the statistical mechanics gives
\begin{equation}
  D_{kl} = \delta_{kl} - \dfrac{V_{k}}{T}
  \left( \ave{\cos kq\cos lq}_{\rm eq} - \ave{\cos kq}_{\rm eq}\ave{\cos lq}_{\rm eq} \right).
\end{equation}
The two $D$ matrices, and $\gamma_{+}$ accordingly,
coincide for homogeneous initial states associated with 
$F_{\rm eq}(\HI) \propto \exp(-p^2/2T)$, 
because $\aveJI{\cos lq}=0$ and $\ave{\cos lq}_{\rm eq} = 0$.

Before progressing to the critical exponents,
we remark on the critical point.
The diagonal elements of $D^{\rm (homo)}$
represent the dispersion functions for the Fourier mode $k$
with the reference state homogeneous as the Para side.
In other words, $D_{kk}^{\rm (homo)}>0$ implies that $m_{k}=0$ is stable.
Assuming that $\FI$ is a monotonically decreasing function of energy,
we have the relation
$V_{k} > V_{l} \Longrightarrow D_{kk}^{\rm (homo)} < D_{ll}^{\rm (homo)}$.
We are focusing on the Para-Ferro phase transition,
and hence $V_{1}$ must be positive
and larger than $V_{2},\cdots,V_{K}$
to make the mode $k=1$ unstable first.
Thus, the critical point is determined by $D_{11}^{\rm (homo)}=0$,
and around it, $D_{11}$ is small but $D_{kk}=O(1)~(2\leq k\leq K)$
in both of the Para and the Ferro sides.
We remark that both the Vlasov dynamics
and the statistical mechanics have the identical critical point,
since they have the identical matrix $D$
in the homogeneous Para side as mentioned above.

Computation of the critical exponent $\beta$ is rather complicated
than the HMF case,
but we can show that the leading term in $\varphi_{1}$ is of $O(m_{1}^{3})$
as the HMF case.
First, we can show that $O(m_{k})\leq O(m_{1}^{2})~(2\leq k\leq K)$,
(see Appendix \ref{sec:order}).
Then, remember that the function $\varphi_{1}(m)$ is obtained
by expanding $\int_{\mu} \FI(\HI) \cos q {\rm d}q{\rm d}p$,
where $m_{k}$ dependence comes from $\HI$ including the term of
$-V_{k}m_{k}\cos kq$.
Thus, terms of $O(m_{1}^{2})$ do not appear in $\varphi_{1}(m)$
since $\int\cos^{3}q{\rm d}q=0$.
On the other hand, terms of $O(m_{1}^{3})$ survive, and by the relation
$O(m_{k})\leq O(m_{1}^{2})$, this is the leading order of $\varphi_{1}$.
Scaling of the spontaneous magnetization is, therefore,
$m_{1}\propto \sqrt{-D_{11}^{\rm (homo)}}$ and $\beta=1/2$ in general.

The linear response for off-critical is obtained by
\begin{equation}
  D ( \Lambda \delta m + h) - h = 0.
\end{equation}
Thus, susceptibility matrix whose $(k,l)$-elements are defined by
\begin{equation}
  \chi_{kl} = \lim_{\norm{h}\to 0} \dfrac{\partial(\delta m_{k})}{\partial h_{l}},
\end{equation}
is expressed as
\begin{equation}
    \chi = \Lambda^{-1} D^{-1} (1-D).
\end{equation}
In the Para side, the off-diagonal elements of $D$
vanish thanks to $\HI=p^{2}/2$ and
$\aveJI{\cos lq}=0$.
The matrix $D$ is hence estimated as
\begin{equation}
 D^{\rm (Para)} =
  {\rm diag}\left(O(D_{11}^{\rm (homo)}), O(1), \cdots O(1)\right)
  .
\end{equation}
This estimation immediately gives $\gamma_{+}=2\beta$ for $\chi_{11}$,
and the other susceptibilities do not diverge.
In the Ferro side, we have
\begin{equation}
  \label{eq:D-Ferro}
  D^{\rm (Ferro)} =
  \begin{pmatrix}
    O(\sqrt{\Delta\HI}) & O(\sqrt{\Delta\HI}) & \cdots & O(\sqrt{\Delta\HI}) \\
    O(\sqrt{\Delta\HI})  & O(1) & \cdots & O(\sqrt{\Delta\HI}) \\
    \vdots & \vdots & \ddots & \vdots \\
    O(\sqrt{\Delta\HI})  & O(\sqrt{\Delta\HI}) & \cdots & O(1) 
  \end{pmatrix}
\end{equation}
with the factor $\Delta\HI=2\sum_{k:{\rm odd}}V_{k}m_{k}$,
which is dominated by $k=1$ from the ordering $O(m_{k})\leq O(m_{1}^{2})$
mentioned previously.
The estimation \eqref{eq:D-Ferro} is obtained as follows.

The ordering also suggests that the fixed point of $\HI$ are solely
$(q,p)=(0,0)$ stable and $(\pi,0)$ unstable,
and therefore, the first diagonal element is estimated
by the same strategy with the HMF case.
Each off-diagonal element is also dominated by $O(\sqrt{\Delta\HI})$
coming from the term having $\aveJI{\cos lq}$,
since the other term gives contribution of higher order $O(m_{1})=O(\Delta\HI)$
from the expansion of $\FI'(\HI)$ with respect to small $m$.

The inverse matrix of $D^{\rm (Ferro)}$ is
\begin{equation}
  \left[ D^{\rm (Ferro)} \right]^{-1} =
  \begin{pmatrix}
    O(1/\sqrt{\Delta\HI}) & O(1)  & \cdots & O(1) \\
    O(1)  & O(1) & \cdots & O(1) \\
    \vdots & \vdots & \ddots & \vdots \\
    O(1)  & O(1) & \cdots & O(1) 
  \end{pmatrix}.
\end{equation}
Therefore, remembering $O(\Delta\HI)=O(m_{1})$
and estimating $[(D^{\rm (Ferro)})^{-1}](1-D^{\rm (Ferro)})$,
the critical exponent for $\chi_{11}$ is $\gamma_{-}=\beta/2$,
and the other elements do not diverge.

The unique divergence in the susceptibility matrix $\chi$ appears in $\chi_{11}$,
and we consider the response to the external force $h=(h_{1},0,\cdots,0)$
at the critical point.
The matrix $D$ does not vanish even at the critical point,
and hence we consider the equation
\begin{equation}
  (D+\Lambda C) (\Lambda\delta m+h) - h = 0.
\end{equation}
The matrix $D+\Lambda C$ can be estimated at the critical point as
$D^{\rm (Ferro)}$, \eqref{eq:D-Ferro},
but replacing $\Delta\HI$ with $\Delta\HA$,
where $\Delta\HA=2\sum_{k:{\rm odd}}(V_{k}\delta m_{k}+h_{k})$.
We may expect that $V_{1}\delta m_{1}+h_{1}$ dominates $\Delta\HA$
and $O(\Delta\HA)=O(V_{1}\delta m_{1}+h_{1})$,
since the susceptibility $\chi_{11}$ diverges at the critical point
but the others do not.
Consequently, we have $(V_{1}\delta m_{1}+h_{1})^{3/2}\propto h_{1}$,
which implies $\delta m_{1} \propto h_{1}^{2/3}$ and $\delta=3/2$.

\section{Summary and discussions}
\label{sec:summary}
We investigated critical exponents
for the continuous Para-Ferro phase transitions in QSSs
from one simple expanded expression of self-consistent equations
for the order parameters like the Landau theory.
The expression is obtained from recently proposed
nonlinear response theory,
and we successfully unified to derive the four critical
exponents $\beta=1/2,~\gamma_{+}=2\beta,~\gamma_{-}=\beta/2$
and $\delta=3/2$ in the HMF model
for reference families of QSSs including thermal equilibrium family.
The unification is further extended into generalized
mean-field 1D systems periodic spatially,
and we obtained the same values for all the critical exponents,
where $\gamma_{\pm}$ and $\delta$ are associated with $\chi_{11}$.
These critical exponents satisfy
the scaling relation $\gamma_{-}=\beta (\delta-1)$.
This relation breaks for $\gamma_{+}$ defined in the Para side,
but it might be reasonable since $\beta$ is defined in the
Ferro side only.
We have also shown that the other elements of susceptibility matrix
do not diverge even at the critical point.

We remark that the essential mechanism of the non-classical
critical exponents, $\gamma_{-}$ and $\delta$,
is existence of averaged factor $\aveJA{\dV}$ 
in \eqref{eq:fAfI-expand}.
Then, from the key scaling of $O(p)=O(\kappa\sqrt{\Delta\HA})$,
the factor gives contribution of order $\sqrt{m_{1}+\delta m_{1}+h_{1}}$
to the self-consistent equations,
and this square root contribution yields
the two non-classical exponents.
We stress that this key scaling sheds light
on understanding the scaling relation $\gamma_{-}=\beta (\delta-1)$
by generalizing the exponent $1/2$ to $x$.
The generalized exponent $x$ gives
$\gamma_{-}=\beta x$ and $\delta=1+x$,
and immediately the scaling relation
irrespective of the value of $\beta$.

We have restricted ourselves to finite $K$,
which is the number of Fourier modes in the interaction $V(q)$.
If we may assume that the interaction $V(q)$ and the external part $H_{\rm ext}(q)$ 
are sufficiently smooth, the amplitudes of their Fourier modes , 
$|V_{k}|$ and $|h_{k}|$, are converges to $0$ rapidly enough.
Then, we conjecture that the higher modes are negligible
and the critical exponents do not change even $K$ is infinite.

The starting equation of the present study corresponds to
the derivative of pseudo free energy in the Landau theory.
Constructing the pseudo free energy might be a future work.
We took averages over iso-action lines
with the aid of an ergodic-like formula \cite{OY14, CL98}.
Spatially higher dimensional systems
are not integrable in general,
but the present theory could be extended by taking the averages
over iso-energy surfaces
if we may use an ergodic-like formula on the surfaces.
The extended theory conserves the Casimirs within the linear order,
and therefore,
can be applied to vaster class of Hamiltonian systems.
We have considered the Hamiltonian external forces
associated with the order parameters,
following the conventional setting of response theory.
Extension to non-Hamiltonian external force or random perturbation
might be other future works.

\begin{acknowledgements}
  The authors thank the anonymous referees for valuable comments
    to improve the manuscript.
  SO is supported by Grant-in-Aid for JSPS Fellows Grant Number 254728.
  YYY acknowledges the support of JSPS KAKENHI Grant Number 23560069.
\end{acknowledgements}

\appendix 

\section{Proof of the ordering $O(m_{k})\leq O(m_{1}^{2})~(k\geq2)$}
\label{sec:order}

We show the ordering of spontaneous order parameters as
$O(m_{k})\leq O(m_{1}^{2})~(k=2,\cdots,K)$
around the critical point of Para-Ferro transition,
which implies $|m_{k}|\ll 1$.
We assume that the function $\FI$ is expanded into the Taylor series.
Using the small $m_{k}$, we expand the self-consistent equation
\begin{equation}
  m_{k} = \int \FI \left(
    p^{2}/2-\sum_{l=1}^{K}V_{l}m_{l}\cos lq \right) \cos kq {\rm d}q{\rm d}p
\end{equation}
as
\begin{equation}
  \begin{split}
    & m_{k} =
    - \sum_{l=1}^{K} V_{l}m_{l} \int \FI'(p^{2}/2) 
    \cos kq \cos lq {\rm d}q{\rm d}p + \cdots. \\
  \end{split}
\end{equation}
We write the expanded equation as
\begin{equation}
  \label{eq:expanded-self-consistent}
  D_{kk}^{\rm (homo)}m_{k} = \varphi_{k}(V_1m_{1},\cdots, V_K m_{K})
\end{equation}
where $\varphi_{k}$ are series 
consisting of monomials whose degrees are more than $1$.
We will derive contradiction by assuming
that there exists $c_{2} \in \{2,\cdots,K\}$
such that $O(m_{c_{2}})> O(m_{1})$.
The contradiction implies $O(m_{c_{2}})\leq O(m_{1})$
for any $c_{2} \in \{2,\cdots,K\}$.
Substituting this relation into \eqref{eq:expanded-self-consistent},
remembering $D_{kk}^{\rm (homo)}=O(1)$
for any $k \geq 2$ in the vicinity of a critical point,
and using that the degree of $\varphi_{k}$ is more than $1$,
we conclude $O(m_{k})=O(\varphi_{k})\leq O(m_{1}^{2})$.

Let us derive the contradiction.
We focus on the equation for $k=c_{2}$.
The left-hand-side of \eqref{eq:expanded-self-consistent}
is of $O(m_{c_{2}})$, and hence
the function $\varphi_{c_{2}}$ must include monomials
of the same order with $m_{c_{2}}$.
We pick up one of them denoted by $\varphi_{c_{2}}^{\ast}$.
Remembering $|m_{l}|\ll 1$ for any $l$
and that the degree of $\varphi_{c_{2}}^{\ast}$ is more than $1$, 
we find that $\varphi_{c_{2}}^{\ast}$ does not include $m_{c_{2}}$.
Next, if $\varphi_{c_{2}}^{\ast}$ includes $m_{1}$,
the same reasoning induces the relation
$O(m_{c_{2}})<O(m_{1})$, but
this breaks the assumption
of $O(m_{c_{2}})>O(m_{1})$. Thus, we conclude that 
$\varphi_{c_{2}}^{\ast}$ includes neither $m_{c_{2}}$ nor $m_{1}$. 
We choose $m_{c_{3}}$ included in $\varphi_{c_{2}}^{\ast}$
such that $c_{3}\in \{2,\cdots,K\}\setminus \{c_{2}\}$
and satisfying $O(m_{1})<O(m_{c_{2}})<O(m_{c_{3}})$,
and we shift the focusing equation to $k=c_{3}$. 
This discussion can repeat up to choosing $c_{K}$,
but no next number $c_{K+1}$ exists.
The nonexistence suggests that there is no monomial in $\varphi_{c_{K}}$
which is of the same order with $m_{c_{K}}$.
The self-consistent equation for $m_{c_{K}}$ is not satisfied,
and a contradiction has been induced. $\blacksquare$

\end{document}